\documentclass[12pt]{article}

\usepackage[totalwidth=460pt,totalheight=600pt]{geometry}
\usepackage{amsfonts,latexsym,amssymb,amsmath,graphicx,accents,slashed,subfigure}
\usepackage[T1]{fontenc}
\usepackage[hidelinks]{hyperref}

\global\arraycolsep=1truept

\numberwithin{equation}{section}

\begin{document}

\bigskip \phantom{C}

\vskip1truecm

\begin{center}
{\huge \textbf{Fakeons And The Classicization}}

\vskip.8truecm

{\huge \textbf{Of Quantum Gravity:}}

\vskip.65truecm

{\huge \textbf{The FLRW Metric}}

\vskip1truecm

\textsl{Damiano Anselmi}

\vskip .1truecm

\textit{Dipartimento di Fisica ``Enrico Fermi'', Universit\`{a} di Pisa}

\textit{and INFN, Sezione di Pisa,}

\textit{Largo B. Pontecorvo 3, 56127 Pisa, Italy}

damiano.anselmi@unipi.it

\vskip2truecm

\textbf{Abstract}
\end{center}

Under certain assumptions, it is possible to make sense of higher derivative
theories by quantizing the unwanted degrees of freedom as fakeons, which are
later projected away. Then the true classical limit is obtained by
classicizing the quantum theory. Since quantum field theory is formulated
perturbatively, the classicization is also perturbative. After deriving a
number of properties in a general setting, we consider the theory of quantum
gravity that emerges from the fakeon idea and study its classicization,
focusing on the FLRW metric. We point out cases where the fakeon projection
can be handled exactly, which include radiation, the vacuum energy density
and the combination of the two, and cases where it cannot, which include
dust. Generically, the classical limit shares many features with the quantum
theory it comes from, including the impossibility to write down complete,
\textquotedblleft exact\textquotedblright\ field equations, to the extent
that asymptotic series and nonperturbative effects come into play.

\vfill\eject

\section{Introduction}

\setcounter{equation}{0}

Typically, higher derivative quantum field theories propagate ghosts, if
they are formulated in the usual ways. The ghosts are unphysical degrees of
freedom that cannot be projected away without violating unitarity. Recently,
a new quantization prescription \cite{LWgrav,fakeons} has been set forth, to
quantize various types of degrees of freedom as \textquotedblleft
fakeons\textquotedblright , i.e. fake particles. The main virtue of the
fakeons is that they can be projected away from the physical spectrum
consistently with unitarity.

The fakeon prescription can be used to turn the ghosts and possibly some
physical particles into fake particles. Its main application is to quantum
gravity \cite{LWgrav,UVQG,absograv}, since one fakeon $\chi _{\mu \nu }$ of
spin two, together with a scalar field $\phi $, is able to make the theory
renormalizable while preserving unitarity.

In this paper we investigate some remarkable features of the classical
limits of the theories of particles and fakeons. We recall that the fakeon
quantization prescription has a truly quantum nature, since it amounts to a
nonanalytic operation on the loop diagrams, called average continuation. The
average continuation is the arithmetic average of the analytic continuations
that circumvent the thresholds associated with the processes that involve
fakeons \cite{LWformulation,fakeons}.

The idea originates from a thorough analysis of the cutting equations, which
are diagrammatic identities that encode the unitarity relation $S^{\dag }S=1$
\cite{cuttingeq}. The fakeons also allow us to reformulate and actually
better understand the Lee-Wick models \cite{LW}. For a review of these
topics, see ref. \cite{classicization}.

\begin{figure}[t]
\begin{center}
\includegraphics[width=16truecm]{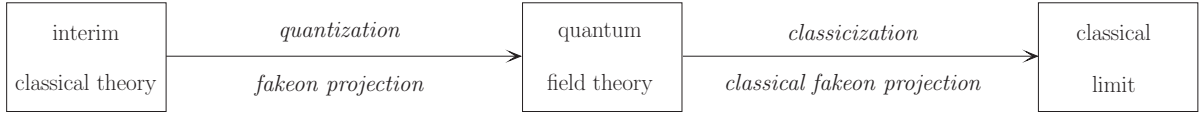}
\end{center}
\caption{Quantization/classicization scheme}
\label{classic}
\end{figure}

The backlash of the fakeon prescription on the classical theory turns out to
be nontrivial \cite{classicization}, because the quantization process
includes an additional step, as shown in fig. \ref{classic}. The starting
local action is just an interim one, being unprojected. The finalized
classical action can be obtained only after the quantization, and emerges
from the classicization of the quantum theory.

The interim classical action of quantum gravity coupled to matter can be
expressed in two ways. The standard way is by means of higher-derivative
terms \cite{LWgrav}: 
\begin{equation}
S_{\text{QG}}(g,\Phi )=-\frac{1}{2\kappa ^{2}}\int \mathrm{d}^{4}x\sqrt{-g}%
\left[ 2\Lambda _{C}+\zeta R+\alpha \left( R_{\mu \nu }R^{\mu \nu }-\frac{1}{%
3}R^{2}\right) -\frac{\xi }{6}R^{2}\right] +S_{\mathfrak{m}}(g,\Phi ).
\label{SQG}
\end{equation}%
Here $\alpha $, $\xi $, $\zeta $ and $\kappa $ are real positive constants.
We make no assumption on the sign of the cosmological constant $\Lambda _{C}$%
. The Planck mass is $M_{\text{Pl}}=1/\sqrt{G}=\sqrt{8\pi \zeta }/\kappa $.
Moreover, $\Phi $ are the matter fields and $S_{\mathfrak{m}}$ is the action
of the matter sector. For example, $S_{\mathfrak{m}}$ can be the action of
the standard model, or a standard model extension, as long as it is
covariantized and contains all the nonminimal couplings that are compatible
with renormalizability. 

For simplicity, in this paper we work at $\Lambda _{C}=0$ and view the
cosmological constant as a component of dark energy. An equivalent version
of the interim classical action (\ref{SQG}) is obtained by means of extra
fields, which allow us to remove the higher derivatives. We find \cite%
{absograv} 
\begin{equation}
\mathcal{S}_{\text{QG}}(g,\phi ,\chi ,\Phi )=S_{\text{H}}(g)+S_{\chi
}(g,\chi )+S_{\phi }(\tilde{g},\phi )+S_{\mathfrak{m}}(\tilde{g}\mathrm{e}%
^{\kappa \phi },\Phi ),  \label{SQG2}
\end{equation}%
where~$\tilde{g}_{\mu \nu }=g_{\mu \nu }+2\chi _{\mu \nu }$ and 
\begin{eqnarray}
&&S_{\text{H}}(g)=-\frac{\zeta }{2\kappa ^{2}}\int \mathrm{d}^{4}x\sqrt{-g}%
R,\qquad S_{\phi }(g,\phi )=\frac{3\zeta }{4}\int \mathrm{d}^{4}x\sqrt{-g}%
\left[ \nabla _{\mu }\phi \nabla ^{\mu }\phi -\frac{m_{\phi }^{2}}{\kappa
^{2}}\left( 1-\mathrm{e}^{\kappa \phi }\right) ^{2}\right] ,  \notag \\
&&S_{\chi }(g,\chi )=S_{\text{H}}(\tilde{g})-S_{\text{H}}(g)+\int \mathrm{d}%
^{4}x\left[ -2\chi _{\mu \nu }\frac{\delta S_{\text{H}}(g)}{\delta g_{\mu
\nu }}+\frac{\zeta ^{2}}{2\alpha \kappa ^{2}}\sqrt{-g}(\chi _{\mu \nu }\chi
^{\mu \nu }-\chi ^{2})\right] _{g\rightarrow \tilde{g}}.  \label{ss}
\end{eqnarray}%
As we see, the theory describes the graviton, a scalar field $\phi $ of
squared mass $m_{\phi }^{2}=\zeta /\xi $, a spin-2 fakeon $\chi _{\mu \nu }$
of squared mass $m_{\chi }^{2}=\zeta /\alpha $ and the matter fields.

It is easy to show, from the expression of $S_{\chi }$, that the $\chi _{\mu
\nu }$ quadratic action is of the Pauli-Fierz type, but with the wrong
overall sign \cite{absograv}. For this reason $\chi _{\mu \nu }$ must be
quantized as a fakeon. At present we do not know whether $\phi $ should be
quantized as a physical particle or a fakeon. Thus, we have two
possibilities: one is the graviton/scalar/fakeon (GSF) theory and the other
one is the graviton/fakeon/fakeon (GFF) theory. Throughout this paper, we
work with the second option, because we plan to investigate the fakeons in
the Friedmann-Lemaitre-Robertson-Walker (FLRW) scenario, which is not
sensitive to $\chi _{\mu \nu }$.

We recall that if we quantize every degree of freedom by means of the
standard Feynman prescription, the action (\ref{SQG}) gives the Stelle
theory \cite{stelle} (after we drop $S_{\mathfrak{m}}$). In that case, no
projection is possible and the classicization is trivial. However, the
Stelle theory propagates ghosts.

The fakeon projection is inherited from quantum field theory, so it is
formulated perturbatively. Its classical limit amounts to take the average
of the retarded and advanced potentials \cite{classicization}. What happens
when we try and resum the perturbative expansion of the classicization? Can
we grasp the \textquotedblleft exact\textquotedblright\ classical field
equations and the fakeon projection at the nonperturbative level? In this
paper, we investigate these issues and uncover interesting, and to some
extent surprising, properties.

At the quantum level, we are accustomed to build a theory perturbatively, by
adding, so to speak, quantum after quantum, or interaction after
interaction. We do not expect anything like that to occur in a classical
framework. One of the surprises of the theory of quantum gravity built on
the fakeon idea is precisely that the classical limit shares many features
with the quantum theory it comes from, including the impossibility to write
down complete, exact\ field equations. Unless we have knowledge about the
nonperturbative sector of quantum gravity, the projected classical field
equations we get are also perturbative. In general, asymptotic series come
into play and nonperturbative corrections may have to be included. However,
in special cases, the resummation can be handled exactly.

We study these issues in a general setting and then concentrate on the FLRW
solution of the classicized theory of quantum gravity. We show that the
fakeon projection can be handled exactly in the cases of radiation, the
vacuum energy density and the combination of both. Instead, in the case of
dust it cannot, so asymptotic series are generated and nonperturbative
effects may come into play.

Quantum gravity, as it emerges from the fakeon idea, is in line with
high-energy particle physics. In particular, it follows from the same
principles that lead to the standard model: unitarity, locality and
renormalizability \cite{QGcorr}. The scattering amplitudes are defined
perturbatively by means of Feynman diagrams, which can be calculated with an
effort comparable to the one required by analogous computations in the
standard model \cite{UVQG,absograv}.

Several proposals for quantum gravity have appeared in the past decades. We
mention string theory \cite{string}, loop quantum gravity \cite{loop},
holography (the AdS/CFT\ correspondence) \cite{ads}, lattice gravity \cite%
{latticeg} and asymptotic safety \cite{asafety}. However, their predictive
powers are limited. Some proposals, like string theory, have a huge space of
free parameters \cite{landscape}. Others, like the AdS/CFT correspondence,
rely on conjectured dualities. Some, like lattice gravity, asymptotic safety
and the AdS/CFT\ correspondence, do not admit perturbative expansions and
deal with strongly coupled quantum field theory. Others, like string theory
and loop quantum gravity, involve mathematics that is not well understood.

Here are some of the reasons why we claim that the solution provided by the
fakeons is the right theory of quantum gravity. As far as calculability,
predictivity and falsifiability are concerned, the fakeon solution tops the
competitors by far. Actually, it may be turn out to be the most predictive
theory ever, since it is able to cover a huge range of energies (from the
infrared limit up to and beyond the Planck scale) perturbatively and with
few independent parameters.

The masses $m_{\phi }$ and $m_{\chi }$ of $\phi $ and $\chi _{\mu \nu }$
might be smaller, or even much smaller, than the Planck mass $M_{\text{Pl}}$%
. The perturbative expansion, which is formulated in powers of the
fakeon/graviton fine structure constants $\alpha _{\phi }=m_{\phi }^{2}/M_{%
\text{Pl}}^{2}$ and $\alpha _{\chi }=m_{\chi }^{2}/M_{\text{Pl}}^{2}$, makes
sense as long as the renormalization group flow keeps these parameters
smaller than unity, which likely means somewhere above the Planck scale. At
some point, up there, nonperturbative effects start to become important. The
theory predicts new physics below the Planck scale \cite{UVQG,absograv}, at
energies around $m_{\phi }$ and $m_{\chi }$. At low energies, it reduces to
the nonrenormalizable theory made of the Hilbert-Einstein action plus the
counterterms turned on by renormalization \cite{newQG}. Note that the
low-energy expansion is independent of the prescription with which the
fields are quantized.

It is important to stress that the fakeon idea does not make assumptions
about the nature of spacetime at infinitesimally small distances. Instead,
the new understanding of spacetime at the microscopic level emerges from the
theory itself. It is encoded in the violation of microcausality \cite%
{absograv,classicization}: the concepts of space and time, past, present and
future, cause and effect lose meaning at energies larger than the lightest
fakeon mass. Our present knowledge of the laws of physics leaves enough room
for this prediction to be accurate, both from the theoretical and
experimental viewpoints.

Over the years, the concept of causality has been gradually put aside in
quantum field theory. The reason is that it is not well understood, which
makes it hard to elevate it to the rank of a fundamental principle. A
definition that matches the intuitive notion is missing \cite{diagrammar}
and Bogoliubov's proposal \cite{bogoliubov}, which implies the
Lehmann-Symanzik-Zimmermann one (i.e. that the fields commute at spacelike
separated points), is an off-shell condition for the Feynman diagrams and
the correlation functions. At the experimental level, the difficulty with
causality comes from the fact it is hard to localize particles described by
relativistic wave packets that are on shell.

The paper is structured as follows. In section \ref{toy} we study the
fakeons and the classicization in nonrelativistic mechanics. In section \ref%
{expansion} we study the asymptotic expansion of the fakeon projection. In
section \ref{issues} we analyze the issues that arise at the nonperturbative
level.\ In section \ref{class} we recall the basic aspects of the
classicization of quantum gravity. In section \ref{redaction} we study the
FLRW\ solution. In section \ref{auxi} we give details on how to proceed in
the non-higher-derivative approach (\ref{SQG2}). Section \ref{conclu}
contains the conclusions.

\section{Fakeon projection in nonrelativistic mechanics}

\setcounter{equation}{0}\label{toy}

In this section and the next one we study the fakeon projection and its
resummation in some models of nonrelativistic mechanics, which provide a
simple environment where most key conceptual issues are already in play. We
consider the higher-derivative Lagrangian 
\begin{equation}
\mathcal{L}_{\text{HD}}=\frac{m}{2}(\dot{x}^{2}-\tau ^{2}\ddot{x}%
^{2})-V(x,t),  \label{lagen2}
\end{equation}%
where $x$ is the coordinate, $m$ is the mass and $\tau $ is a real constant.

The simplest case is $V(x,t)=-xF_{\text{ext}}(t)$, where $F_{\text{ext}}(t)$
is an external force. The unprojected equation of motion is $mK\ddot{x}=F_{%
\text{ext}}$, where%
\begin{equation}
K=1+\tau ^{2}\frac{\mathrm{d}^{2}}{\mathrm{d}t^{2}},  \label{H}
\end{equation}%
and the projected one reads 
\begin{equation}
m\ddot{x}=\left\langle F_{\text{ext}}\right\rangle _{K}.  \label{average}
\end{equation}%
As recalled in the introduction, the classical fakeon average is 
\begin{equation}
\langle A\rangle _{X}\equiv \frac{1}{2}\left[ \left. \frac{1}{X}\right\vert
_{\text{rit}}+\left. \frac{1}{X}\right\vert _{\text{adv}}\right] A,
\label{fave}
\end{equation}%
the subscripts denoting the retarded and advanced potentials, respectively.
We find \cite{classicization} 
\begin{equation}
m\ddot{x}=\int_{-\infty }^{\infty }\mathrm{d}u\hspace{0in}\hspace{0.01in}%
\frac{\sin (|u|/\tau )}{2\tau }F_{\text{ext}}(t-u).  \label{preq}
\end{equation}

\subsection{Fakeon averages}

Before moving to the cases where the resummation of the fakeon projection
plays an important role, it is useful to check out the fakeon average $%
\langle F_{\text{ext}}\rangle _{K}$ in some simple examples. If the external
force is Gaussian, 
\begin{equation}
F_{\text{ext}}(t)=\exp \left( -\frac{\gamma }{2}t^{2}\right) ,  \label{gau}
\end{equation}%
the fakeon average returns a wiggling function, as shown in fig. \ref%
{gausian}:%
\begin{equation*}
\langle F_{\text{ext}}\rangle _{K}=\sqrt{\frac{\pi }{2\gamma }}\mathrm{e}%
^{-1/(2\gamma )}\,\mathrm{Im}\left[ \mathrm{e}^{it}\text{Erf}\left( \frac{%
\gamma t+i}{\sqrt{2\gamma }}\right) \right] .
\end{equation*}%
\begin{figure}[t]
\begin{center}
\includegraphics[width=16truecm]{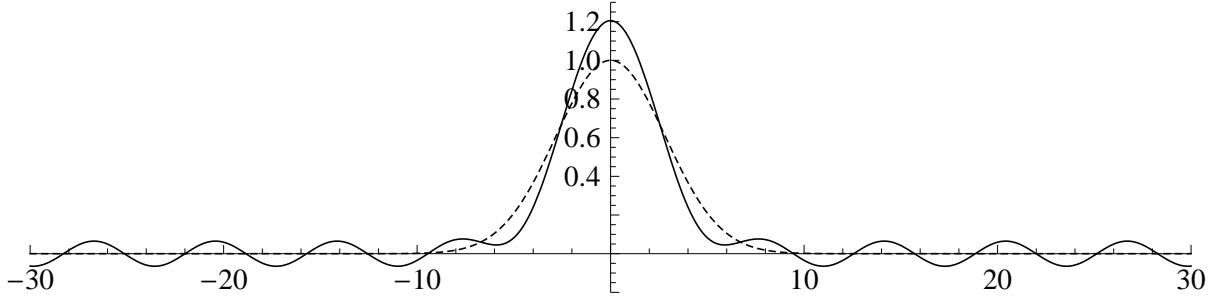}
\end{center}
\caption{Fakeon average (continuous line) of a Gaussian function (\protect
\ref{gau}) (dashed line) with $\protect\tau =1$ and $\protect\gamma =1/8$.}
\label{gausian}
\end{figure}

The average $\langle P_{n}(t)\rangle _{K}$ of a polynomial $P_{n}(t)$ of
degree $n$ is another polynomial $Q_{n}(t)$ of the same degree, which can be
determined from $KQ_{n}=P_{n}$. For example, $\langle 1\rangle _{K}=1$, $%
\langle t\rangle _{K}=t$, $\langle t^{2}\rangle _{K}=t^{2}-2\tau ^{2}$.
These results can be also verified by taking the limits%
\begin{equation}
\lim_{\gamma \rightarrow 0}\left\langle P_{n}(t)\mathrm{e}^{-\gamma
t^{2}/2}\right\rangle _{K}.  \label{limi}
\end{equation}%
Similarly, we find%
\begin{equation*}
\langle \mathrm{e}^{i\omega t}\rangle _{K}=\frac{\mathrm{e}^{i\omega t}}{%
1-\omega ^{2}\tau ^{2}},\qquad \langle \mathrm{e}^{it/\tau }\rangle _{K}=%
\frac{\mathrm{e}^{it/\tau }}{4\tau }(\tau -2it),
\end{equation*}%
etc., for $\omega <1/\tau $.

The resummation of the fakeon projection often leads to multiple averages,
such as $\langle \langle F_{\text{ext}}\rangle \rangle $, $\langle \langle
\langle F_{\text{ext}}\rangle \rangle \rangle $, etc. If we want to know how
to handle these expressions, we must go back to the origin of the
projection, rooted in quantum field theory. In ref. \cite{fakeons} it was
shown that when two or more fakeon thresholds coincide, they must be treated
as limits of distinct thresholds. From this property we can easily prove the
identity 
\begin{equation}
\lim_{\epsilon \rightarrow 0}\mathcal{P}\prod\nolimits_{i=1}^{n+1}\frac{1}{%
x-\epsilon c_{i}}=\frac{(-1)^{n}}{n!}\frac{\mathrm{d}^{n}}{\mathrm{d}x^{n}}%
\mathcal{P}\frac{1}{x},  \label{theo}
\end{equation}%
where $\mathcal{P}$ denotes the principal value and $c_{i}$ are arbitrary
distinct numbers. This formula allows us to \textquotedblleft raise\ $%
\mathcal{P}$ to arbitrary powers\textquotedblright\ and so compute the
multiple averages.

Specifically, if $\tilde{F}_{\text{ext}}(\nu )$ is the Fourier transform of $%
F_{\text{ext}}(t)$, we have%
\begin{equation*}
\langle F_{\text{ext}}\rangle _{K}=\mathcal{P}\int_{-\infty }^{+\infty }%
\frac{\mathrm{d}\nu }{2\pi }\frac{\mathrm{e}^{-i\nu t}\tilde{F}_{\text{ext}%
}(\nu )}{1-\tau ^{2}\nu ^{2}}.
\end{equation*}%
\textquotedblleft Squaring the average\textquotedblright\ by means of (\ref%
{theo}), we find 
\begin{eqnarray}
\langle \langle F_{\text{ext}}\rangle _{K}\rangle _{K} &=&\langle F_{\text{%
ext}}\rangle _{K}+\frac{1}{2}\frac{\mathrm{d}}{\mathrm{d}t}\left[ \langle
tF_{\text{ext}}\rangle _{K}-t\langle F_{\text{ext}}\rangle _{K}\right] 
\notag \\
&=&\int_{-\infty }^{\infty }\hspace{0in}\frac{\mathrm{d}u}{4}\left( \sin
|u|-|u|\cos u\right) F_{\text{ext}}(t-\tau u).  \label{squa}
\end{eqnarray}%
With the help of a limit like (\ref{limi}), it is easy to check that $%
\langle \langle 1\rangle _{K}\rangle _{K}=1$, $\langle \langle t\rangle
_{K}\rangle _{K}=t$ and $\langle \langle t^{2}\rangle _{K}\rangle
_{K}=t^{2}-4\tau ^{2}$. In analogous ways, formulas for more repeated
averages can be worked out.

\subsection{Harmonic oscillator with an external force}

The resummation of the projection is important in the next example, which is
the harmonic oscillator with an external force: 
\begin{equation*}
V(x,t)=\frac{m}{2}\omega ^{2}x^{2}-xF_{\text{ext}}(t).
\end{equation*}%
We view $\omega ^{2}$ as the expansion parameter. The unprojected equation
of motion is%
\begin{equation}
mK\ddot{x}+m\omega ^{2}x=F_{\text{ext}}=m\tilde{K}\left( \frac{\mathrm{d}^{2}%
}{\mathrm{d}t^{2}}+\Omega ^{2}\right) x,  \label{resunpro}
\end{equation}%
where%
\begin{equation*}
\Omega =\frac{1}{\tau \sqrt{2}}\sqrt{1-\sqrt{1-4\tau ^{2}\omega ^{2}}}%
,\qquad \tilde{\Omega}=\frac{1}{\tau \sqrt{2}}\sqrt{1+\sqrt{1-4\tau
^{2}\omega ^{2}}},\qquad \tilde{K}=\tau ^{2}\tilde{\Omega}^{2}+\tau ^{2}%
\frac{\mathrm{d}^{2}}{\mathrm{d}t^{2}}.
\end{equation*}%
The resummed projected equation, which makes sense for $\omega <1/(2\tau )$,
can be quickly obtained by inverting the operator $\tilde{K}$ according to
the classical fakeon prescription. The result is 
\begin{equation}
m\left( \frac{\mathrm{d}^{2}}{\mathrm{d}t^{2}}+\Omega ^{2}\right) x=\langle
F_{\text{ext}}\rangle _{\tilde{K}}=\int_{-\infty }^{\infty }\mathrm{d}u%
\hspace{0in}\hspace{0.01in}\hspace{0.01in}\frac{\sin \left( \tilde{\Omega}%
|u|\right) }{2\tau ^{2}\tilde{\Omega}}\hspace{0.01in}F_{\text{ext}}(t-u).
\label{respro}
\end{equation}%
If we expand the average back in powers of $\omega ^{2}$, we find 
\begin{equation*}
\langle F_{\text{ext}}\rangle _{\tilde{K}}=\langle F_{\text{ext}}\rangle
_{K}+\tau ^{2}\omega ^{2}(1+\tau ^{2}\omega ^{2})\langle \langle F_{\text{ext%
}}\rangle _{K}\rangle _{K}+\tau ^{4}\omega ^{4}\langle \langle \langle F_{%
\text{ext}}\rangle _{K}\rangle _{K}\rangle _{K}+\mathcal{O}(\omega ^{6}),
\end{equation*}%
which shows that the identity (\ref{theo}) is crucial to deal with the
multiple averages that lead to the projected equation (\ref{respro}) from
the unprojected equation (\ref{resunpro}).

The fakeons that are projected away are the solutions of $\tilde{K}x=0$,
i.e. 
\begin{equation*}
x(t)=C\cos \left( \tilde{\Omega}t+\varphi \right) .
\end{equation*}

The\ result of the resummation highlights some nontrivial, nonperturbative
effects that come into play beyond the convergence radius of the expansion.
Indeed, for $\omega >1/(2\tau )$ the frequencies $\Omega $ and $\tilde{\Omega%
}$ become complex and the fakeon projection jumps into another
\textquotedblleft phase\textquotedblright , where all four independent
solutions are unacceptable and must be projected away.

In more complicated cases it may be hard to tell what the fakeon projection
becomes nonperturbatively. In principle, settling this issue requires
knowledge of the nonperturbative sector of quantum field theory. However,
workarounds are available in lucky situations, as we show in section \ref%
{redaction}.

\section{Fakeon projection by asymptotic expansion}

\setcounter{equation}{0}\label{expansion}

When the potential $V$ contains anharmonic terms, the equations must be
treated self consistently. One way to handle the fakeon projection, which we
investigate in this section, is by means of an iterative procedure. The
projected equations that we obtain are nonpolynomial and must in general be
interpreted as asymptotic expansions.

To begin with, let us consider the Lagrangian (\ref{lagen2}) with the
potential 
\begin{equation}
V=\frac{m}{2}\omega ^{2}x^{2}+\frac{\lambda }{4!}x^{4}.  \label{quartic}
\end{equation}%
The unprojected equation of motion is 
\begin{equation}
m\left( \frac{\mathrm{d}^{2}}{\mathrm{d}t^{2}}+\tau ^{2}\frac{\mathrm{d}^{4}%
}{\mathrm{d}t^{4}}+\omega ^{2}\right) x=m\tilde{K}\left( \frac{\mathrm{d}^{2}%
}{\mathrm{d}t^{2}}+\Omega ^{2}\right) x=-\frac{\lambda x^{3}}{3!}.
\label{unpro}
\end{equation}%
We assume $\omega <1/(2\tau )$. If we resum the expansion in powers of $%
\omega ^{2}$ as explained in the previous section, we obtain the projected
equation%
\begin{equation}
m\left( \frac{\mathrm{d}^{2}}{\mathrm{d}t^{2}}+\Omega ^{2}\right) x=-\frac{%
\lambda }{3!}\langle x^{3}\rangle _{\tilde{K}},  \label{proquart}
\end{equation}%
which must still be understood perturbatively in $\lambda $.

One way to deal with (\ref{proquart}) is to search for a solution of the form%
\begin{equation*}
x(t)=x_{0}(t)+\sum_{n=1}^{\infty }\tilde{\lambda}^{n}x_{n}(t),
\end{equation*}%
where $\tilde{\lambda}=\lambda /m$ and $x_{0}(t)$ solves the homogeneous
equation $\ddot{x}_{0}=-\Omega ^{2}x_{0}$. We get%
\begin{equation*}
\left( \frac{\mathrm{d}^{2}}{\mathrm{d}t^{2}}+\Omega ^{2}\right) x_{1}=-%
\frac{1}{3!}\langle x_{0}^{3}\rangle _{\tilde{K}},\qquad \left( \frac{%
\mathrm{d}^{2}}{\mathrm{d}t^{2}}+\Omega ^{2}\right) x_{2}=-\frac{1}{2}%
\langle x_{1}x_{0}^{2}\rangle _{\tilde{K}},
\end{equation*}%
etc., which can be solved by means of the fakeon averages and the rules
outlined before.

Another way is to write a generic expansion for the right-hand side, 
\begin{equation}
\left( \frac{\mathrm{d}^{2}}{\mathrm{d}t^{2}}+\Omega ^{2}\right)
x=x\sum_{n=1}^{\infty }\tilde{\lambda}^{n}\tau
^{2n-2}\sum_{k=0}^{n}c_{n,k}x^{2n-2k}(\tau \dot{x})^{2k},  \label{expa}
\end{equation}%
insert it into the unprojected equation (\ref{unpro}) and determine the
unknown coefficients $c_{n,k}$ by matching the monomials. So doing, we can
build the projected equation to arbitrarily high orders in $\tilde{\lambda}$%
. To the first order, we obtain%
\begin{equation*}
\left( \frac{\mathrm{d}^{2}}{\mathrm{d}t^{2}}+\Omega ^{2}\right) x=-\frac{%
\tilde{\lambda}x\left[ (\tilde{\Omega}^{2}-7\Omega ^{2})x^{2}-6\dot{x}^{2}%
\right] }{6\tau ^{2}(\tilde{\Omega}^{2}-\Omega ^{2})(\tilde{\Omega}%
^{2}-9\Omega ^{2})}+\mathcal{O}(\tilde{\lambda}^{2}).
\end{equation*}%
At higher orders we find very involved expressions. For the sake of
simplicity, from this point onwards we take $\omega =0$ (which means $\Omega
=0$, $\tilde{\Omega}=1/\tau $). Every result can be generalized
straightforwardly to nonvanishing $\omega $. To the third order we obtain 
\begin{eqnarray}
\ddot{x} &=&-\frac{\tilde{\lambda}x}{6}\left( x^{2}-6\tau ^{2}\dot{x}%
^{2}\right) -\frac{\tilde{\lambda}^{2}\tau ^{2}x}{12}\left( x^{4}-48\tau
^{2}x^{2}\dot{x}^{2}+372\tau ^{4}\dot{x}^{4}\right)  \notag \\
&&-\frac{\tilde{\lambda}^{3}\tau ^{4}x}{6}\left( x^{6}-156\tau ^{2}x^{4}\dot{%
x}^{2}+4572\tau ^{4}x^{2}\dot{x}^{4}-31152\tau ^{6}\dot{x}^{6}\right) +%
\mathcal{O}(\tilde{\lambda}^{4}).  \label{xdotdot}
\end{eqnarray}

The truncation of the projected equation to a finite order $n$ in $\tilde{%
\lambda}$ is polynomial. The expansion is asymptotic and the coefficients
grow very fast, although slower than $(4n)!$. In this table we give the
orders of magnitude of the coefficients $c_{n,0}$ and $c_{n,n}$ for various
values of $n$, which we have computed up to $n=25$: 
\begin{equation}
\begin{tabular}{|c|c|c|c|c|c}
\hline
$n$ & $5$ & $10$ & $15$ & $20$ & $25$ \\ \hline
$c_{n,0}$ & $10^{0}$ & $10^{6}$ & $10^{13}$ & $10^{22}$ & $10^{32}$ \\ \hline
$c_{n,n}$ & $10^{9}$ & $10^{28}$ & $10^{52}$ & $10^{78}$ & $10^{107}$%
\end{tabular}
\label{table}
\end{equation}

Note that the expansion we are dealing with does not coincide with the
\textquotedblleft low-energy\textquotedblright\ expansion in powers of $\tau
^{2}$, which treats the higher-derivative term $\tau ^{2}\mathrm{d}^{2}/%
\mathrm{d}t^{2}$ as small. Instead, we are expanding in the dimensionless
parameter $\tilde{\lambda}$, so each truncation gives a solution that in
principle holds for all times. The price we pay is that we have to handle
more involved truncations. Indeed, the coefficient $c_{n,k}$, which is
multiplied by $\tilde{\lambda}^{n}\tau ^{2n+2k-2}$, becomes relevant at the $%
n$th order of the expansion in powers of $\tilde{\lambda}$, but only at the $%
(n+k+1)$th order of the expansion in powers of $\tau ^{2}$. As we see
already from (\ref{xdotdot}), the latter grows much more slowly than the
former.

The expansion of the projected Lagrangian $\mathcal{L}$ can be worked out in
a similar way. We write a generic expansion in $x$, $\dot{x}$ and determine
its coefficients by demanding that the Lagrange equations be equivalent to (%
\ref{xdotdot}). The result is 
\begin{equation*}
\frac{\mathcal{L}}{m}=\frac{\dot{x}^{2}}{2}-\frac{\tilde{\lambda}x^{2}}{4!}%
\left( x^{2}+12\tau ^{2}\dot{x}^{2}\right) +\frac{\tau ^{2}\tilde{\lambda}%
^{2}x^{2}}{72}(x^{4}-54\tau ^{2}x^{2}\dot{x}^{2}+372\tau ^{4}\dot{x}^{4})+%
\mathcal{O}(\tilde{\lambda}^{3}).
\end{equation*}%
The energy $E$ can be obtained from $\mathcal{L}$ or, again, by writing the
most general expansion and working out the coefficients that make $\mathrm{d}%
E/\mathrm{d}t$ vanish on the solutions of (\ref{xdotdot}): 
\begin{equation*}
\frac{E}{m}=\frac{\dot{x}^{2}}{2}+\frac{\tilde{\lambda}x^{2}}{4!}\left(
x^{2}-12\tau ^{2}\dot{x}^{2}\right) -\frac{\tau ^{2}\tilde{\lambda}^{2}x^{2}%
}{72}(x^{4}+54\tau ^{2}x^{2}\dot{x}^{2}-1116\tau ^{4}\dot{x}^{4})+\mathcal{O}%
(\tilde{\lambda}^{3}).
\end{equation*}

For each truncation to order $n$, the projected equations can be solved
numerically. If we compare the solutions for growing $n$, we observe the
typical behaviors of the asymptotic solutions. The lowest values of $n$ give
results that are acceptable, but not very accurate. Then we find stable
results in a certain window $n_{1}\leqslant n\leqslant n_{2}$, which
provides the best approximation of the exact solution. Finally, unreliable
behaviors appear for $n>n_{2}$. On general grounds, $n_{2}$ is proportional
to $1/\tilde{\lambda}$. Asymptotic expansions cannot be arbitrarily precise,
but in several situations the window $n_{1}\leqslant n\leqslant n_{2}$ is
precise enough.

\begin{figure}[t]
\begin{center}
\includegraphics[width=16truecm]{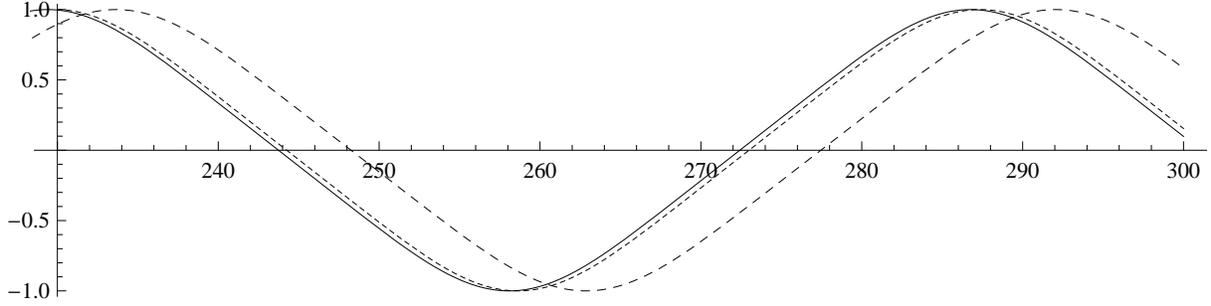}
\end{center}
\caption{Solution $x(t)$ of the truncated equation (\protect\ref{xdotdot})
for $x(0)=1$, $\dot{x}(0)=0$, $m=\protect\tau =1$, $\protect\lambda =1/10$.
The sparsely dashed line is $n=1$. The densely dashed line is $n=2$, while
the continuous line is $n=3$. The solution remains stable from $n=3$ to $%
n=10 $.}
\label{NonAsy}
\end{figure}

For example, with the initial conditions $x(0)=1$, $\dot{x}(0)=0$ and $%
m=\tau =1$, $\lambda =1/10$, we find the trajectories of fig. \ref{NonAsy}.
The solution with $n=1$ is not very accurate, while the one with $n=2$ is
considerably better. The trajectory remains stable in the window $3\leqslant
n\leqslant 10$. The robust stability is a benefit of the stability of the
potential (\ref{quartic}). With different values of $\lambda $ we find $%
n_{2}\sim 1/\lambda $. 

\section{The nonperturbative fakeon projection}

\setcounter{equation}{0}\label{issues}

The nonperturbative fakeon projection can only follow from the knowledge of
the nonperturbative sector of quantum field theory. Indeed, it is easy to
show that, when asymptotic expansions are the best we have, at the
nonperturbative level the arbitrariness associated with the essential
singularities takes us back to the unprojected equations.

Let $S_{\text{HD}}(\phi ,\lambda ,\tau )$ denote a higher-derivative action
that depends on the fields $\phi ^{i}$, $i=1,\ldots N$, and their first $M$
time derivatives. Let $\lambda $ denote the couplings, such that $S_{\text{HD%
}}(\phi ,0,\tau )$ is free. Let $\tau $ denote the parameters that multiply
the higher-derivative corrections, such that $S_{\text{HD}}(\phi ,\lambda
,0) $ is the non-higher-derivative action.

We assume that all the degrees of freedom due to the higher derivatives are
quantized as fakeons. We focus on the dependence on the time coordinate $t$
and ignore any dependence on the space coordinates $x$, $y$, $z$. It is
understood that, when we talk about initial or integration \textquotedblleft
constants\textquotedblright , they may be functions of $x$, $y$ and $z$. We
also assume that the fields $\phi ^{i}$ are \textquotedblleft
bosonic\textquotedblright , so the field equations depend on $\phi ^{i}$, $%
\dot{\phi}^{i}$ and $\ddot{\phi}^{i}$ at $\tau =0$.

We have three versions of the classical equations:

($a$) the higher-derivative equations%
\begin{equation}
E_{\text{HD}}^{i}(\phi ,\lambda ,\tau )=0,  \label{eqd}
\end{equation}%
which are exact, but unprojected; they are satisfied by the acceptable
solutions, but also by the fakeon solutions, which must be discarded;

($b$) the projected equations%
\begin{equation}
E_{\text{P}}^{i}(\phi ,\lambda ,\tau )=0  \label{ep}
\end{equation}%
which are understood perturbatively in $\lambda $;

($b$) the exact projected equations%
\begin{equation}
E_{\text{PnP}}^{i}(\phi ,\lambda ,\tau )=0,  \label{exep}
\end{equation}%
which can in principle be determined by studying the nonperturbative sector
of the parent quantum field theory.

In the example treated above, ($a$) are (\ref{unpro}) and ($b$) are (\ref%
{proquart}). We assume that ($c$) are not known. However, we assume that
they exist.

Now, let%
\begin{equation}
\phi ^{i}=f^{i}(t,\lambda ,\tau ,c^{ia})  \label{soleq}
\end{equation}%
denote the solutions of (\ref{eqd}), where $c^{ia}$ are the integration
constants ($a=1,\ldots M$) that parametrize the initial conditions. The
solutions of the exact projected equations (\ref{exep}) are particular cases
of (\ref{soleq}). They have the form%
\begin{equation}
\phi ^{i}=f^{i}(t,\lambda ,\tau ,d^{ia}(a^{i},b^{i},\lambda ,\tau )),
\label{nonpp}
\end{equation}%
where the constants $d^{ia}$ are not independent, but functions%
\begin{equation}
d^{ia}(a^{i},b^{i},\lambda ,\tau )  \label{cia}
\end{equation}%
of $\lambda $, $\tau $ and $2N$ independent integration constants $a^{i}$, $%
b^{i}$. The solutions of (\ref{ep}) are particular cases of (\ref{nonpp}), 
\begin{equation}
\phi ^{i}=f^{i}(t,\lambda ,\tau ,c^{ia}(a^{i},b^{i},\lambda ,\tau )),
\label{psol}
\end{equation}%
where the functions $c^{ia}(a^{i},b^{i},\lambda ,\tau )$ are only known as
asymptotic expansions in powers of $\lambda $ and coincide with the
asymptotic expansions of (\ref{cia}).

The difference $d^{ia}(a^{i},b^{i},\lambda ,\tau
)-c^{ia}(a^{i},b^{i},\lambda ,\tau )$ is made of essential singularities for 
$\lambda \rightarrow 0$, which cannot be worked out from the sole knowledge
of ($a$) and ($b$). If we attempt a resummation (with the Borel method, for
example, when applicable), the solution will unlikely satisfy (\ref{eqd}).
The space of functions that have the same asymptotic expansions and satisfy (%
\ref{eqd}) at the same time obviously coincides with the space of
unprojected solutions (\ref{soleq}).

This means that, unless we have direct knowledge about the nonperturbative
sector of the parent quantum field theory, we cannot write \textquotedblleft
exact\textquotedblright\ classical field equations and mostly have to work
with their perturbative form.

However, workarounds may be available in special cases by means of
resummations. Even in quantum field theory we have example of exact results
that can be derived form the perturbative expansion. We mention the
anomalies (which are one-loop exact), the renormalization group flow (which
allows us to resum the leading logs, the next-to-leading logs, etc.), the
particle self-energies, obtained by resumming the bubble diagrams (which
give the particle lifetimes, among other things), and so on. Similarly,
there are cases where, in spite of the difficulties stressed in this
section, we can get to the exact projected solutions (\ref{nonpp}) in
quantum gravity. In the following sections we describe some important
examples.

\section{The classical limit of quantum gravity}

\setcounter{equation}{0}\label{class}

Before proceeding, we briefly recall the basic aspects of the classicization
of quantum gravity. At the conceptual level, it is convenient to work with
the non-higher-derivative interim classical action (\ref{SQG2}). The field
equations of the metric read%
\begin{equation}
R^{\mu \nu }-\frac{1}{2}g^{\mu \nu }R=\frac{\kappa ^{2}}{\zeta }\left[ 
\mathrm{e}^{3\kappa \phi }fT_{\mathfrak{m}}^{\mu \nu }(\tilde{g}\mathrm{e}%
^{\kappa \phi },\Phi )+fT_{\phi }^{\mu \nu }(\tilde{g},\phi )+T_{\chi }^{\mu
\nu }(g,\chi )\right] ,  \label{meq}
\end{equation}%
where $T_{A}^{\mu \nu }(g)=-(2/\sqrt{-g})(\delta S_{A}(g)/\delta g_{\mu \nu
})$ are the energy-momentum tensors ($A=\mathfrak{m}$, $\phi $, $\chi $) and 
$f=\sqrt{\det \tilde{g}_{\rho \sigma }/\det g_{\alpha \beta }}$. The field
equations of the fakeons $\phi $ and $\chi _{\mu \nu }$ are \cite%
{classicization}%
\begin{eqnarray}
-\frac{1}{\sqrt{-\tilde{g}}} &&\partial _{\mu }\left( \sqrt{-\tilde{g}}%
\tilde{g}^{\mu \nu }\partial _{\nu }\phi \right) -\frac{m_{\phi }^{2}}{%
\kappa }\left( \mathrm{e}^{\kappa \phi }-1\right) \mathrm{e}^{\kappa \phi }=%
\frac{\kappa \mathrm{e}^{3\kappa \phi }}{3\zeta }T_{\mathfrak{m}}^{\mu \nu }(%
\tilde{g}\mathrm{e}^{\kappa \phi },\Phi )\tilde{g}_{\mu \nu },  \notag \\
&&\frac{1}{\sqrt{-g}}\frac{\delta S_{\chi }(g,\chi )}{\delta \chi _{\mu \nu }%
}=\mathrm{e}^{3\kappa \phi }fT_{\mathfrak{m}}^{\mu \nu }(\tilde{g}\mathrm{e}%
^{\kappa \phi },\Phi )+fT_{\phi }^{\mu \nu }(\tilde{g},\phi ).  \label{feq}
\end{eqnarray}

Let $\langle \phi \rangle $ and $\langle \chi _{\mu \nu }\rangle $ denote
the solutions of the equations (\ref{feq}), obtained with the half sum of
the retarded and advanced Green functions. The projected field equations are
(\ref{meq}), once $\phi $ and $\chi _{\mu \nu }$ are replaced by $\langle
\phi \rangle $ and $\langle \chi _{\mu \nu }\rangle $. They can also be
derived as Lagrange equations of the finalized classical action 
\begin{equation}
\mathcal{S}_{\text{QG}}^{\text{GFF}}(g,\Phi )=S_{\text{H}}(g)+S_{\chi
}(g,\langle \chi \rangle )+S_{\phi }(\bar{g},\langle \phi \rangle )+S_{%
\mathfrak{m}}(\bar{g}\mathrm{e}^{\kappa \langle \phi \rangle },\Phi ).
\label{sgff}
\end{equation}%
where~$\bar{g}_{\mu \nu }=g_{\mu \nu }+2\langle \chi _{\mu \nu }\rangle $.

As said, we have to understand the projection perturbatively and deal with
the issues explained in the previous sections. In the next sections we study
the resummation of the perturbative projection in the case of the FLRW
solution. At the practical level, it is more convenient to work with the
interim action (\ref{SQG}), but in section \ref{auxi} we give details on how
to obtain the same results by working with (\ref{SQG2}).

\section{The classicization of the FLRW solution}

\setcounter{equation}{0}\label{redaction}

It is often convenient to search for solutions of the field equations
starting from an ansatz, as in the case of the FLRW metric. However, in
general, it is not legitimate to insert the ansatz directly into the action
and work out the Lagrange equations of the so-obtained reduced action.
Indeed, the ansatz reduces the space of configurations. A minimum, or more
generally extremum, of the action on the reduced space of configurations is
not guaranteed to be a minimum or extremum on the full space.

However, under certain conditions it is possible to obtain the correct
equations of motion by applying the variational principle to the reduced
action. We derive the key properties to achieve this goal and then apply the
method of the reduced action to the FLRW ansatz.

\subsection{Method of the reduced action}

Consider an action $S(\phi )$ depending on the fields $\phi ^{i}$, $%
i=1,\ldots N$. The Lagrange equations are 
\begin{equation}
\frac{\delta S}{\delta \phi ^{i}}=0.  \label{unred}
\end{equation}%
Consider an ansatz 
\begin{equation}
\phi ^{i}=f^{i}(\varphi )  \label{ansatz}
\end{equation}%
that expresses the fields $\phi ^{i}$ in terms of a reduced set of fields $%
\varphi ^{\alpha }$, $\alpha =1,\ldots M$, with $M<N$. The reduced action is
then 
\begin{equation*}
S_{r}(\varphi )=S(f(\varphi ))
\end{equation*}%
and its field equations read 
\begin{equation}
0=\frac{\delta S_{r}(\varphi )}{\delta \varphi ^{\alpha }}=\left. \frac{%
\delta S}{\delta \phi ^{i}}\right\vert _{\phi =f(\varphi )}\frac{\delta
f^{i}(\varphi )}{\delta \varphi ^{\alpha }}.  \label{red}
\end{equation}%
Now, assume that

($i$) the $M$ equations\ (\ref{red}) are independent, and

($ii$) $M$ equations (\ref{unred}) are independent and the other $N-M$
equations (\ref{unred}) are algebraic relations among the $M$ independent
ones.

Then, the equations (\ref{unred}) are equivalent to the equations (\ref{red}%
) derived from the reduced action $S_{r}$.

Typically, point ($ii$) can be established by means of symmetry arguments
and other properties of the ansatz. Point ($i$) is easy to check directly.

If the relations mentioned in point ($ii$) happen to be differential instead
of algebraic, further assumptions must be advocated to obtain the right set
of equations after the reduction.

\subsection{The FLRW\ metric}

Now we apply the method of the reduced action to the FLRW metric, which we
parametrize as 
\begin{equation}
\mathrm{d}s^{2}=g_{\mu \nu }\mathrm{d}x^{\mu }\mathrm{d}x^{\nu }=b^{2}(t)%
\mathrm{d}t^{2}-a^{2}(t)\mathrm{d}\sigma ^{2},  \label{flrw}
\end{equation}%
where, in spherical polar coordinates, 
\begin{equation*}
\mathrm{d}\sigma ^{2}=\frac{\mathrm{d}r^{2}}{1-kr^{2}}+r^{2}\mathrm{d}\theta
^{2}+r^{2}\sin ^{2}\theta \hspace{0.01in}\mathrm{d}\phi ^{2}.
\end{equation*}%
The lapse function $b(t)$ is inserted to meet the requirements explained
above and keep track of the time reparametrizations. Indeed, we know that
the FLRW\ ansatz reduces the field equations to two independent ones, so we
need two independent functions $a(t)$ and $b(t)$ to have a meaningful
reduced action $S_{r}$. We can set $b(t)\equiv $ $1$ after applying the
variational principle to $S_{r}$.

Under the usual assumptions of homogeneity and isotropy, the matter stress
tensor is 
\begin{equation}
(T_{\mathfrak{m}})_{\mu }^{\nu }=\rho (t)\delta _{0}^{\nu }\delta _{\mu
}^{0}-p(t)\delta _{i}^{\nu }\delta _{\mu }^{i},  \label{stress}
\end{equation}%
where $\rho $ is the energy density and $p$ is the pressure, $i=1,2,3$ being
a space index. Then the reduced version of the action (\ref{SQG}) of quantum
gravity coupled to matter reads 
\begin{equation}
S_{\text{QG}}\rightarrow -\frac{1}{16\pi G}\frac{r^{2}\sin \theta }{\sqrt{%
1-kr^{2}}}\int \mathrm{d}t\hspace{0.01in}a^{3}b\mathcal{R}\left( 1-\frac{%
\mathcal{R}}{6m_{\phi }^{2}}\right) +S_{\mathfrak{m}},  \label{sarrow}
\end{equation}%
where $m_{\phi }^{2}=\zeta /\xi $ and the Ricci curvature for the ansatz (%
\ref{flrw}) is 
\begin{equation}
\mathcal{R}=-6\left( \frac{\ddot{a}}{ab^{2}}+\frac{\dot{a}^{2}}{a^{2}b^{2}}-%
\frac{\dot{a}\dot{b}}{ab^{3}}+\frac{k}{a^{2}}\right) .  \label{curlyr}
\end{equation}%
The arrow in formula (\ref{sarrow})\ and in the formulas below means that we
ignore the integrals on $r$ and the angles $\theta $ and $\varphi $, which
give an overall (infinite) factor that can be dropped for the purpose of
applying the variational principle.

Note that the $\alpha $-dependent terms of (\ref{SQG}) cancel out, because
they are proportional to the square of the Weyl tensor $C^{\mu \nu \rho
\sigma }$, up to a total derivative, and $C^{\mu \nu \rho \sigma }$ vanishes
identically for the metric (\ref{flrw}).

We do not have a well-defined expression for $S_{\mathfrak{m}}$, with the
stress tensor (\ref{stress}). However, the infinitesimal variation $\delta
S_{\mathfrak{m}}$ is enough for our purposes. It reads 
\begin{equation*}
\delta S_{\mathfrak{m}}=-\frac{1}{2}\int \mathrm{d}^{4}x\hspace{0.01in}\sqrt{%
-g}(T_{\mathfrak{m}})_{\mu }^{\nu }g^{\mu \rho }\delta g_{\nu \rho
}\rightarrow \frac{r^{2}\sin \theta }{\sqrt{1-kr^{2}}}\int \mathrm{d}t%
\hspace{0.01in}a^{2}\left( 3pb\delta a-\rho a\delta b\right) .
\end{equation*}

If we vary the reduced action with respect to $a$ and $b$ and then set $%
b(t)\equiv $ $1$, we obtain the unprojected equations 
\begin{equation}
\Sigma \left( \frac{\ddot{a}}{a}+\frac{\dot{a}^{2}}{a^{2}}+\frac{k}{a^{2}}%
\right) =\frac{4\pi G}{3}(\rho -3p),\qquad \Upsilon \left( \frac{\ddot{a}}{a}%
-\frac{\dot{a}^{2}}{a^{2}}-\frac{k}{a^{2}}\right) =-4\pi G(\rho +p),
\label{roz}
\end{equation}%
where $\Sigma $ and $\Upsilon $ are the operators 
\begin{equation}
\Sigma =1+\frac{1}{m_{\phi }^{2}}\left( 3\frac{\dot{a}}{a}+\frac{\mathrm{d}}{%
\mathrm{d}t}\right) \frac{\mathrm{d}}{\mathrm{d}t},\qquad \Upsilon =\Sigma +%
\frac{2}{m_{\phi }^{2}}\left[ \frac{k}{a^{2}}+3\frac{\mathrm{d}}{\mathrm{d}t}%
\left( \frac{\dot{a}}{a}\right) \right] .  \label{unp}
\end{equation}

The continuity equation 
\begin{equation}
\dot{\rho}+3(\rho +p)\frac{\dot{a}}{a}=0  \label{conse}
\end{equation}%
is the same as usual. It follows from the conservation of the stress-energy
tensor and can be checked by solving (\ref{roz}) for $\rho $ and $p$.

It is easy to verify that the equations (\ref{roz}) match those obtained by
inserting the ansatz (\ref{flrw}) with $b(t)\equiv $ $1$ into the field
equations of (\ref{SQG}) (which can be found for example in ref. \cite%
{classicization}), as guaranteed by the method of the reduced action.

\subsection{Projection}

Since the left-hand sides of the equations (\ref{roz}) factorize the
operators $\Sigma $ and $\Upsilon $, the resummed fakeon projection is
straightforward. If we multiply (\ref{roz}) by $\Sigma ^{-1}$ and $\Upsilon
^{-1}$, defined by means of the classical fakeon prescription, we obtain the
projected equations 
\begin{eqnarray}
\frac{\ddot{a}}{a}+\frac{\dot{a}^{2}}{a^{2}}+\frac{k}{a^{2}} &=&\frac{4\pi G%
}{3}\langle \rho -3p\rangle _{\Sigma },  \label{eq1} \\
\frac{\ddot{a}}{a}-\frac{\dot{a}^{2}}{a^{2}}-\frac{k}{a^{2}} &=&-4\pi
G\langle \rho +p\rangle _{\Upsilon },  \label{eq2}
\end{eqnarray}%
where the fakeon averages are defined in (\ref{fave}).

For some purposes, it is convenient to define a modified energy density $%
\tilde{\rho}$ and a modified pressure $\tilde{p}$ as 
\begin{eqnarray}
\tilde{\rho} &=&\frac{1}{4}\langle \rho -3p\rangle _{\Sigma }+\frac{3}{4}%
\langle \rho +p\rangle _{\Upsilon },  \label{rot} \\
\tilde{p} &=&\frac{1}{4}\langle \rho +p\rangle _{\Upsilon }-\frac{1}{4}%
\langle \rho -3p\rangle _{\Sigma },  \label{pt}
\end{eqnarray}%
and rearrange (\ref{eq1}) and (\ref{eq2}) in forms that match the usual
Friedmann equations: 
\begin{eqnarray}
\frac{\dot{a}^{2}}{a^{2}}+\frac{k}{a^{2}} &=&\frac{8\pi G}{3}\tilde{\rho},
\label{Eq1} \\
2\frac{\ddot{a}}{a}+\frac{\dot{a}^{2}}{a^{2}}+\frac{k}{a^{2}} &=&-8\pi G%
\tilde{p}.  \label{Eq2}
\end{eqnarray}

Adding the derivative of (\ref{Eq1}) to a suitable linear combination of the
two equations, it is easy to get the second continuity equation 
\begin{equation}
\frac{\mathrm{d}\tilde{\rho}}{\mathrm{d}t}+3\frac{\dot{a}}{a}(\tilde{\rho}+%
\tilde{p})=0,  \label{Conse}
\end{equation}%
satisfied by the modified energy density and pressure.

Depending on the problem at hand, the fakeon projection encoded in the
equations (\ref{eq1}) and (\ref{eq2}) may or may not be the final, exact
one. It is exact in some important cases, which include the vacuum energy
density, radiation and their combination. It is not exact in other cases,
which include dust (cold matter). There, however, approximate solutions are
enough for most purposes.

\subsubsection*{Vacuum energy density}

Now we show that in the case of the vacuum energy the solutions of the
projected equations coincide with the solutions of the Friedmann equations
that follow from Einstein gravity.

The equation of state is $p=-\rho $, so the continuity equation (\ref{conse}%
) gives $\rho =\rho _{0}=$ constant. It is convenient to start by solving (%
\ref{eq2}), since its right-hand side vanishes. The solution reads 
\begin{equation}
a(t)=\mathrm{e}^{\sigma t}+\frac{k}{4\sigma ^{2}}\mathrm{e}^{-\sigma t},
\label{coccosta}
\end{equation}%
where $\sigma $ is another constant. The third integration constant has been
absorbed into a time translation.

To study (\ref{eq1}), note that (\ref{coccosta}) and $\rho =\rho _{0}$ imply 
$\Sigma \rho =\rho $. Thus, we also have $\langle \rho \rangle _{\Sigma
}=\rho $ and, from (\ref{rot})-(\ref{pt}), $\tilde{\rho}=\rho $, $\tilde{p}%
=p=-\rho $. Then, equation (\ref{eq1}) gives a relation between the two
constants $\rho _{0}$ and $\sigma $, which reads%
\begin{equation}
\rho _{0}=\frac{3\sigma ^{2}}{8\pi G}.  \label{rocosta}
\end{equation}

\subsubsection*{Radiation}

Similar conclusions hold in the case of radiation, where $p=\rho /3$. The
continuity equation (\ref{conse}) gives 
\begin{equation}
\rho (t)=\frac{\rho _{0}^{\prime }}{a^{4}},  \label{rorad}
\end{equation}%
where $\rho _{0}^{\prime }$ is constant. Solving (\ref{eq1}), whose
right-hand side vanishes, we get 
\begin{equation}
a(t)=\sqrt{t(\sigma ^{\prime }-kt)},  \label{arad}
\end{equation}%
up to a time translation, $\sigma ^{\prime }$ being another constant.

Using (\ref{rorad}) and (\ref{arad}) we easily find $\Upsilon \rho =\rho $,
so $\langle \rho \rangle _{\Upsilon }=\rho $, $\tilde{\rho}=\rho $, $\tilde{p%
}=p=\rho /3$. Then, equation (\ref{eq2}) gives 
\begin{equation*}
\rho _{0}^{\prime }=\frac{3\sigma ^{\prime \hspace{0.01in}2}}{32\pi G}.
\end{equation*}

\subsubsection*{Combination of radiation and vacuum energy density}

Consider the equation of state%
\begin{equation}
p=\frac{\rho }{3}+p_{0}=\frac{1}{3}(\rho -4\rho _{0}),  \label{mix}
\end{equation}%
where $\rho _{0}$ and $p_{0}=-4\rho _{0}/3$ are constants. The interesting
feature of (\ref{mix}) is that it allows us to treat the combination of
radiation and the vacuum energy density, which can be useful to study
inflation. As before, we can solve the projected equations exactly, since (%
\ref{rot}) and (\ref{pt}) give $\tilde{p}=(\tilde{\rho}-4\rho _{0})/3$. For
convenience, we write $\rho _{0}=3\sigma ^{2}/(8\pi G)$. The continuity
equation (\ref{Conse}) gives%
\begin{equation}
\tilde{\rho}(t)=\frac{3}{8\pi G}\left( \sigma ^{2}+\frac{\sigma ^{\prime 
\hspace{0.01in}2}}{4a^{4}}\right) ,  \label{mixro}
\end{equation}%
where $\sigma ^{\prime }$ is constant. Inserting this solution into (\ref%
{Eq1}), we get%
\begin{equation}
a(t)=\sqrt{\frac{\sinh (\sigma t)}{\sigma }\left( \sigma ^{\prime }\cosh
(\sigma t)-\frac{k}{\sigma }\sinh (\sigma t)\right) },  \label{mixa}
\end{equation}%
up to a time translation. We can check that (\ref{mixa}) satisfies (\ref{Eq2}%
) identically. When $\sigma \rightarrow 0$ we retrieve (\ref{arad}). For $%
\sigma ^{\prime }\rightarrow 0$ and $k<0$, we obtain a time-translated
version of (\ref{coccosta}).

We can find the energy density $\rho $ from the second unprojected equation
of formula (\ref{roz}). The result is very similar to (\ref{mixro}),%
\begin{equation*}
\rho (t)=\frac{3}{8\pi G}\left( \sigma ^{2}+\frac{\sigma ^{\prime \prime 
\hspace{0.01in}2}}{4a^{4}}\right) ,\qquad \sigma ^{\prime \prime \hspace{%
0.01in}2}=\sigma ^{\prime 2}\left( 1+\frac{4\sigma ^{2}}{m_{\phi }^{2}}%
\right) ,
\end{equation*}%
the only change being the coefficient of the contribution due to the
radiation. This is also the only correction to the result obtained from the
Einstein equations.

\subsubsection*{General case}

In general, if we assume the equation of state $p=w\rho $, the modified
pressure and density satisfy an $a$-dependent differential equation of
state, which reads 
\begin{equation}
\tilde{p}=w\tilde{\rho}-\frac{(1+w)(1-3w)}{3(1+w)\Sigma +(1-3w)\Upsilon }%
\Delta \tilde{\rho}.  \label{ptw}
\end{equation}%
where $\Delta =\Upsilon -\Sigma $ and the reciprocal operator that appears
here has to be defined by means of the fakeon prescription (\ref{fave}).

The continuity equation (\ref{conse}) gives the usual relation%
\begin{equation*}
\rho (t)=\frac{3\sigma ^{\prime \hspace{0.01in}2}}{32\pi G}\frac{1}{%
a^{3(1+w)}},
\end{equation*}%
where $\sigma ^{\prime }$ is constant. It is convenient to introduce a
function $u(t)$ by writing%
\begin{equation*}
a(t)=\left[ 3\sigma ^{\prime }(1+w)u(t)/4\right] ^{2/(3(1+w))}.
\end{equation*}%
Then the unprojected equations (\ref{roz}) give, in the simple case $k=0$,%
\begin{equation}
m_{\phi }^{2}(1-\dot{u}^{2})=2\dot{u}\dddot{u}-\ddot{u}^{2}-\frac{4w}{1+w}%
\frac{\dot{u}^{2}\ddot{u}}{u}-\frac{1-3w}{1+w}\frac{\dot{u}^{4}}{u^{2}}.
\label{unj}
\end{equation}

The fakeon projection of this equation is rather hard, since it contains no
parameter that we can use to approach the problem perturbatively, other than 
$\tau \equiv 1/m_{\phi }$. If we expand in powers of $\tau $ we obtain the
usual low-energy expansion
\begin{equation}
u(t)=t\left[ 1-\frac{1-3w}{2(1+w)}\frac{1}{m_{\phi }^{2}t^{2}}+\frac{%
13+34w-219w^{2}}{24(1+w)^{2}}\frac{1}{m_{\phi }^{4}t^{4}}+\mathcal{O}\left( 
\frac{1}{m_{\phi }^{6}t^{6}}\right) \right] .  \label{ut}
\end{equation}%
As in section \ref{expansion}, the series is asymptotic and the numerical
coefficients grow very fast. For example, in the case of dust ($w=0$) the
coefficient of $1/(m_{\phi }t)^{14}$ is of order $10^{8}$. Depending on the
values of $m_{\phi }t$, various terms of the asymptotic expansion may offer
a stable, satisfactory approximation of the exact solution.

The reason why the projections that appear in (\ref{eq1}) and (\ref{eq2}),
which are resummed versions of those obtained by expanding around flat
space, are not the exact projections for the problem we are dealing with, is
that the solutions for $a(t)$, $\rho (t)$ and $p(t)$ must be worked out
self-consistently. The equations would be exact if $a(t)$ had to be found
for given $\rho (t)$ and $p(t)$ (see comments below).

We expect that the masses of the fakeons $\chi _{\mu \nu }$ and (possibly) $%
\phi $ have values that are much smaller than the Planck mass. On general
grounds, they could be around $10^{12}$GeV \cite{absograv}. If that is the
case, the value of the parameter $\tau \sim 1/m_{\chi }\sim 1/m_{\phi }$,
which multiplies the higher time derivative $\mathrm{d}/\mathrm{d}t$, is
around $10^{-36}$s. We know that the first moments of the life of the
universe were dominated by radiation, with a crossover to matter dominance
at $t\sim 5\cdot 10^{4}$ years. The matter dominated epoch lasted about $%
10^{10}$ years, followed by the dark energy era. Thus, an exact treatment of
the matter dominated epoch is not strictly necessary in cosmology and the
first few orders of (\ref{ut}) can be enough for most purposes. As shown
previously, the radiation dominated era can be treated exactly, even in
superposition with the vacuum energy density.

\subsubsection*{Other cases where the projection can be worked out exactly}

We conclude by pointing out other situations where the projection can be
handled exactly. The first case is when we need to find the FLRW metric for
given sources, i.e. $\rho $ and $p$ do not have to be determined
self-consistently, but are given functions, known from the start. Then, the
projection encoded in equations (\ref{eq1}) and (\ref{eq2}) is exact. The
solutions do not coincide with those predicted by Einstein gravity and
averages similar to those found in section \ref{toy} appear. The case is to
some extent similar to the case of the harmonic oscillator with an external
force, whose fakeon projection is encoded in formula (\ref{respro}).

We stress that, on the contrary, when $\rho $, $p$ have to be solved
self-consistently together with the metric, the projection contained in the
equations (\ref{eq1}) and (\ref{eq2}) must still be understood
perturbatively. The iterative methods of section \ref{expansion} can be used
to work out the asymptotic expansions of the solutions, which may be
satisfactory for some purposes. An example is the FLRW metric for
nonrelativistic matter.

The second case where we can handle the fakeon projection exactly is when
for some reason we are given an equation of state expressing $\tilde{p}$ as
a function of $\tilde{\rho}$ only. Then, the problem of solving the
equations (\ref{Eq1}) and (\ref{Eq2}), with the help of (\ref{Conse}),
matches the problem of solving the Friedmann equations of Einstein gravity.

One may wonder whether it is possible to make the fakeon averages
effectively disappear by redefining the density and pressure everywhere, so
that $\tilde{\rho}$ and $\tilde{p}$ describe the quantities we really
observe or measure, instead of $\rho $ and $p$. In general, it is not
legitimate to do so, but in some cases, depending on the data available to
us, we may have no other option. More precisely, the relations (\ref{rot})
and (\ref{pt}) between $\tilde{\rho}$, $\tilde{p}$ and $\rho $, $p$ depend
on the particular problem we deal with, to the extent that they contain the
metric and the ansatz we are using. Other problems may lead to different
formulas for the modified quantities $\tilde{\rho}$ and $\tilde{p}$.
Moreover, different interactions, such as the electromagnetic ones, are
sensitive to the unmodified $\rho $ and $p$. Thus, it possible to probe the
relations between $\tilde{\rho}$, $\tilde{p}$ and $\rho $, $p$ by comparing
different physical situations. However, when these comparisons are out of
reach, maybe because not enough data are available, it may be impossible to
tell that equations (\ref{Eq1}) and (\ref{Eq2}) are actually descendants of
the parent equations (\ref{eq1}) and (\ref{eq2}).

A similar conclusion extends to the problem of detecting the violations of
microcausality. Unless we are able to cross check different physical
situations, it may be impossible to uncover the violation, because it may be
easily hidden inside redefinitions of the quantities we measure.


\section{Non-higher derivative approach to the FLRW\ solution}

\setcounter{equation}{0}\label{auxi}

For completeness, we report how the solutions are worked out from the action
(\ref{SQG2}). We start from the ansatz 
\begin{eqnarray}
g_{\mu \nu }\mathrm{d}x^{\mu }\mathrm{d}x^{\nu } &=&\bar{b}^{2}(t)\mathrm{d}%
t^{2}-\bar{a}^{2}(t)\mathrm{d}\sigma ^{2},  \notag \\
\chi _{\mu \nu }\mathrm{d}x^{\mu }\mathrm{d}x^{\nu } &=&d(t)\mathrm{d}%
t^{2}-e(t)\mathrm{d}\sigma ^{2},\qquad \phi =\phi (t).  \label{ans2}
\end{eqnarray}
With this choice, we have the right amount of independent functions to
derive the field equations by means of the reduced action approach.
Alternatively, we can insert the ansatz directly into (\ref{meq}) and (\ref%
{feq}).

Anticipating the result, it is convenient to define%
\begin{equation*}
\bar{b}^{2}=B^{2}-2d,\qquad \bar{a}^{2}=A^{2}-2e,\qquad A=a\mathrm{e}%
^{-\kappa \phi /2},\qquad B=b\mathrm{e}^{-\kappa \phi /2}.
\end{equation*}%
The metric that effectively couples to matter reads%
\begin{equation*}
\tilde{g}_{\mu \nu }\mathrm{e}^{\kappa \phi }\mathrm{d}x^{\mu }\mathrm{d}%
x^{\nu }=b^{2}(t)\mathrm{d}t^{2}-a^{2}(t)\mathrm{d}\sigma ^{2},
\end{equation*}%
where $\tilde{g}_{\mu \nu }=g_{\mu \nu }+2\chi _{\mu \nu }$.

Now we study the $\bar{a}$, $\bar{b}$, $\phi $, $d$ and $e$ field equations,
starting from the $\phi $ one, which reads 
\begin{equation}
\Sigma (1-\mathrm{e}^{-\kappa \phi })=-\frac{8\pi G}{3m_{\phi }^{2}}(\rho
-3p).  \label{phieq}
\end{equation}%
It can be projected straightforwardly, leading to 
\begin{equation}
1-\mathrm{e}^{-\kappa \phi }=-\frac{8\pi G}{3m_{\phi }^{2}}\langle \rho
-3p\rangle _{\Sigma }.  \label{fieq}
\end{equation}%
If we set 
\begin{equation}
\mathrm{e}^{-\kappa \phi }=1-\frac{\mathcal{R}}{3m_{\phi }^{2}},
\label{fiset}
\end{equation}%
where $\mathcal{R}$ is still given by (\ref{curlyr}), equation (\ref{fieq})
becomes equivalent to equation (\ref{eq1}).

Since the FLRW metric has a vanishing Weyl tensor, the functions $d(t)$ and $%
e(t)$ should make the $\alpha $ dependence disappear from the field
equations. This goal is achieved by choosing 
\begin{equation*}
d(t)=\frac{1}{m_{\chi }^{2}}\left( \frac{\dot{A}^{2}}{A^{2}}-2\frac{\ddot{A}%
}{A}+2\frac{\dot{A}\dot{B}}{AB}+k\frac{B^{2}}{A^{2}}\right) ,\qquad e(t)=-%
\frac{1}{m_{\chi }^{2}}\left( \frac{\dot{A}^{2}}{B^{2}}+k\right) ,
\end{equation*}%
where $m_{\chi }^{2}=\zeta /\alpha $. Once we set $b(t)\equiv 1$, we can
drop the $\chi _{\mu \nu }$ field equations, obtained from the variations
with respect to $d$ and $e$, since it is easy to prove that they are
equivalent to the equations obtained from $\bar{a}$ and $\bar{b}$. At the
end, the equations (\ref{meq}) coincide with (\ref{roz}) and can be
projected as before, leading to (\ref{eq1}) and (\ref{eq2}).

In the case of radiation, (\ref{fiset}) gives $\phi =0$, while in the case
of the vacuum energy density we obtain 
\begin{equation}
\phi =-\frac{1}{\kappa }\ln \left( 1+\frac{4\sigma ^{2}}{m_{\phi }^{2}}%
\right) ,  \label{fid}
\end{equation}%
where $\sigma $ is the constant appearing in (\ref{coccosta}). Formula (\ref%
{fid}) also holds in the case of radiation combined with the vacuum energy
density. Conversely, if we start from the ansatz $\phi =$ constant, equation
(\ref{phieq}) implies $\rho -3p=$ constant, which is the equation of state
of the combination of radiation and the vacuum energy density.

We see that by extending the standard FLRW\ ansatz (\ref{flrw}) to (\ref%
{ans2}), the presence of the $\chi _{\mu \nu }$ does not affect the
solution. It is conceivable that many results obtained in inflationary
cosmology \cite{linde} can be extended to the full theory of quantum gravity
studied here, which has the advantage of being renormalizable.

\section{Conclusions}

\setcounter{equation}{0}\label{conclu}

When fakeons are present, the starting, local classical action is just an
interim one. The true classical action emerges only at the very end, after
the quantization, by means of a process of classicization of the quantum
theory. The reason is that the fakeon prescription is not classical, but
emerges from the loop corrections.

Quantum field theory is formulated perturbatively, so the classicization is
also perturbative. The consequences of this fact are quite striking: instead
of having complete, exact classical equations, we deal with the typical
problems of quantum field theory, even if we work at the classical level.
These include the appearance of asymptotic series (when we write the
equations, not just when we search for their solutions) and possibly
important roles played by the nonperturbative corrections. As far as we
know, this backlash of the quantization on the classical limit is
unprecedented.

We have investigated the problems related to the resummation of the
perturbative expansion associated with the fakeon projection and applied the
results to the FLRW metric in quantum gravity. In some cases (like the
vacuum energy, radiation and the combination of the two), the fakeon
projection can be resummed to all orders. In more general cases, which
include dust, asymptotic series are generated and nonperturbative effects
may come into play. The implications on the very early stages of the big
bang remain to be explored.

\section*{Acknowledgments}

We are grateful to U. Aglietti, L. Bracci, M. Piva and T. Morris for useful
discussions.

\end{document}